\documentclass[a4paper,11pt]{article}
\usepackage{pos}

\title{Novel machine learning applications at the LHC}

\author*{Javier M. Duarte}
\onbehalf{on behalf of the ALICE, ATLAS, CMS, and LHCb Collaborations}

\affiliation{Department of Physics\\
University of California San Diego\\
  9500 Gilman Drive\\
  La Jolla, CA 92093}

\emailAdd{jduarte@ucsd.edu}

\abstract{
Machine learning (ML) is a rapidly growing area of research in the field of particle physics, with a vast array of applications at the CERN LHC.
ML has changed the way particle physicists conduct searches and measurements as a versatile tool used to improve existing approaches and enable fundamentally new ones.
In these proceedings, we describe novel ML techniques and recent results for improved classification, fast simulation, unfolding, and anomaly detection in LHC experiments.
}

\FullConference{42nd International Conference on High Energy Physics (ICHEP2024)\\
18-24 July 2024\\
Prague, Czech Republic\\}

\begin{document}
\maketitle

\section{Introduction}

\vspace{-.5em}

 Particle physicists have a long history of developing and applying machine learning (ML) techniques.
From early applications of neural networks to charged particle tracking in the 1980s~\cite{Denby:1987rk} to the Higgs boson discovery in 2012, in which boosted decision trees improved the sensitivity to the $H\to \tau\tau$ decay mode~\cite{Radovic:2018dip}, ML has changed the way particle physicists conduct searches and measurements.
It is an essential and versatile tool that we use to improve existing approaches, and it enables fundamentally new approaches.
In recent years, the subfield of ML in particle physics has grown exponentially in the number of publications and expanded to cover a wide variety of topics and use cases, as indexed by the HEP ML Living Review~\cite{hep_ml_living_review}.% and visualized in Fig.~\ref{fig:nomological_net}.

In these proceedings, we present selected recent results that highlight how LHC experiments are applying novel ML techniques.
In particular, we briefly describe the ML techniques and results for improved classification, faster simulation, unfolding, and anomaly detection.

%\begin{figure}
%    \centering
%    \includegraphics[width=0.5\linewidth]{figures/nomological_net.pdf}
%    \caption{Nomological net of topics in ML in particle physics inspired by the HEP ML Living Review~\cite{hep_ml_living_review}.}
%    \label{fig:nomological_net}
%\end{figure}

\vspace{-.5em}

\section{Improved classification}

\vspace{-.5em}

\begin{figure}[htpb]
    \centering
    \includegraphics[width=0.5\linewidth]{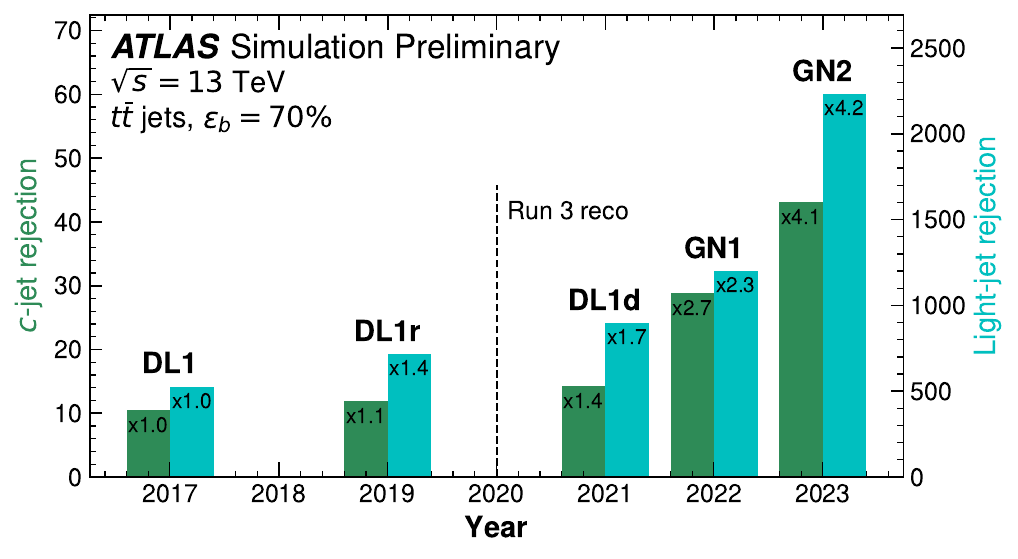}
    \includegraphics[width=0.29\linewidth]{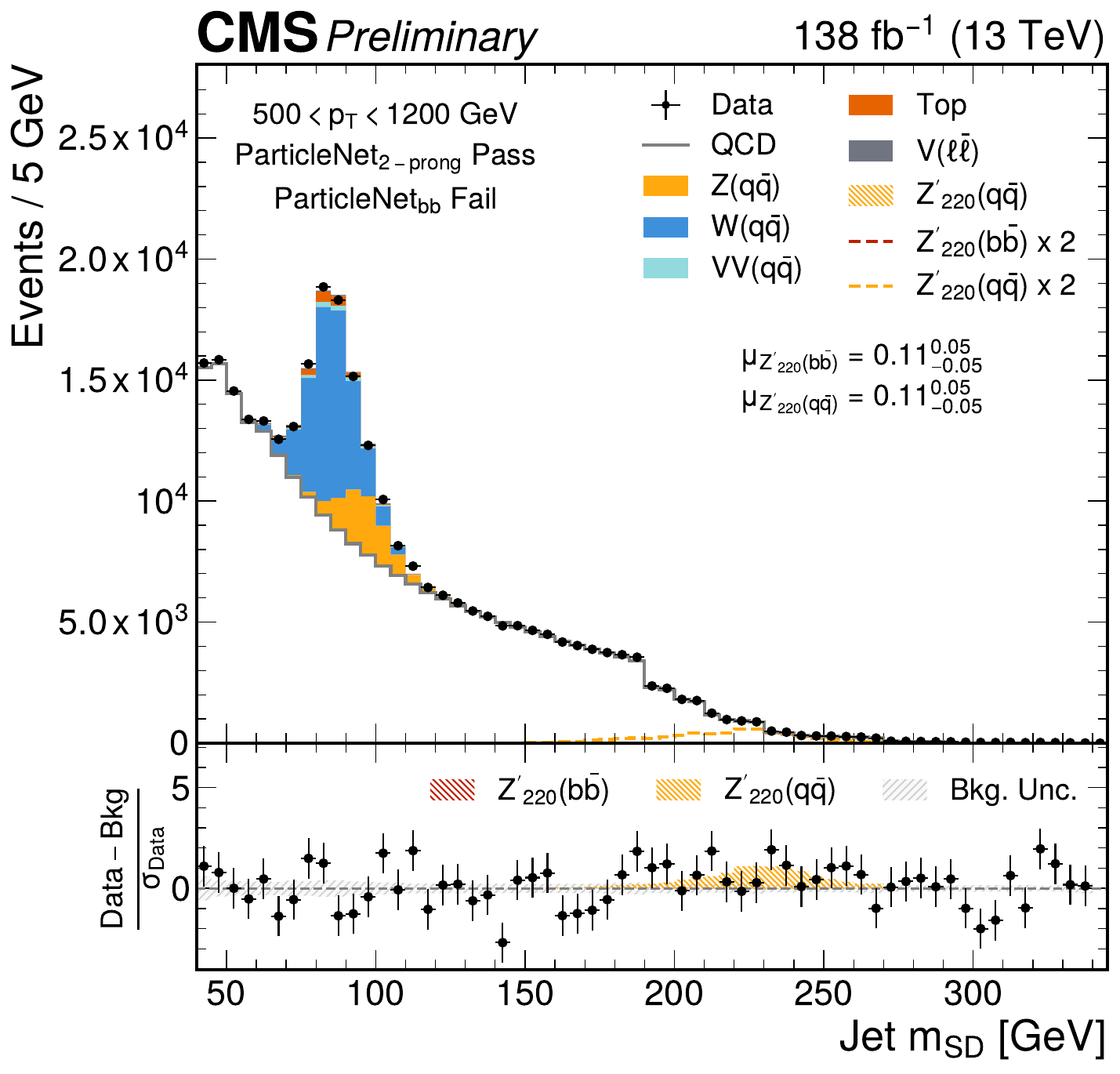}
    \caption{%Visualization of different jet representations, including high-level features, sequences, images, and graphs (upper).
    The $c$ and light jet rejection of the different ATLAS flavor tagging algorithms over time in Monte Carlo simulation (left)~\cite{ATLAS-PLOTS-FTAG-2023-01}.
    The jet soft drop mass distribution after two-prong and light-flavor ParticleNet tagger selections, illustrating the prominent $W(qq)$ resonance (right)~\cite{CMS-PAS-EXO-24-007}.
    }
    \label{fig:ATLASCMSGNNs}
\end{figure}

Great strides have been made to leverage rich low-level information for a variety of tasks, including jet classification, by considering different \emph{representations}.
Traditionally, jets are preprocessed into a vector of high-level features for use with ML methods like feedforward neural networks or boosted decision trees.
More recently, jets have been represented as sequences, images, or graphs; the latter can be processed by graph neural networks~\cite{Moreno:2019bmu} or attention-based transformers~\cite{vaswani2017attention}.
For example, major improvements have been demonstrated in CMS with ParticleNet~\cite{Qu:2019gqs} and Particle Transformer (ParT)~\cite{Qu:2022mxj} and in ATLAS with the GN1~\cite{ATL-PHYS-PUB-2022-027} and GN2~\cite{ATLAS-PLOTS-FTAG-2023-01} heavy flavor taggers.
The improvement over the years of increasingly sophisticated $b$-tagging algorithms in ATLAS is shown in Fig.~\ref{fig:ATLASCMSGNNs} (left).
In particular, the latest GN2 algorithm boasts a factor of 4.2 improvement in light jet rejection over the baseline.
Figure~\ref{fig:ATLASCMSGNNs} (right) also shows the power of the CMS ParticleNet $X(qq)$ large-radius jet tagger in a search for light $Z'$ resonances~\cite{CMS-PAS-EXO-24-007}.
The tagger is able to reduce the substantial QCD multijet background to reveal the prominent $W(qq)$ resonance in the jet soft drop mass spectrum.

\begin{figure}[htpb]
    \centering
    \includegraphics[width=0.3\linewidth]{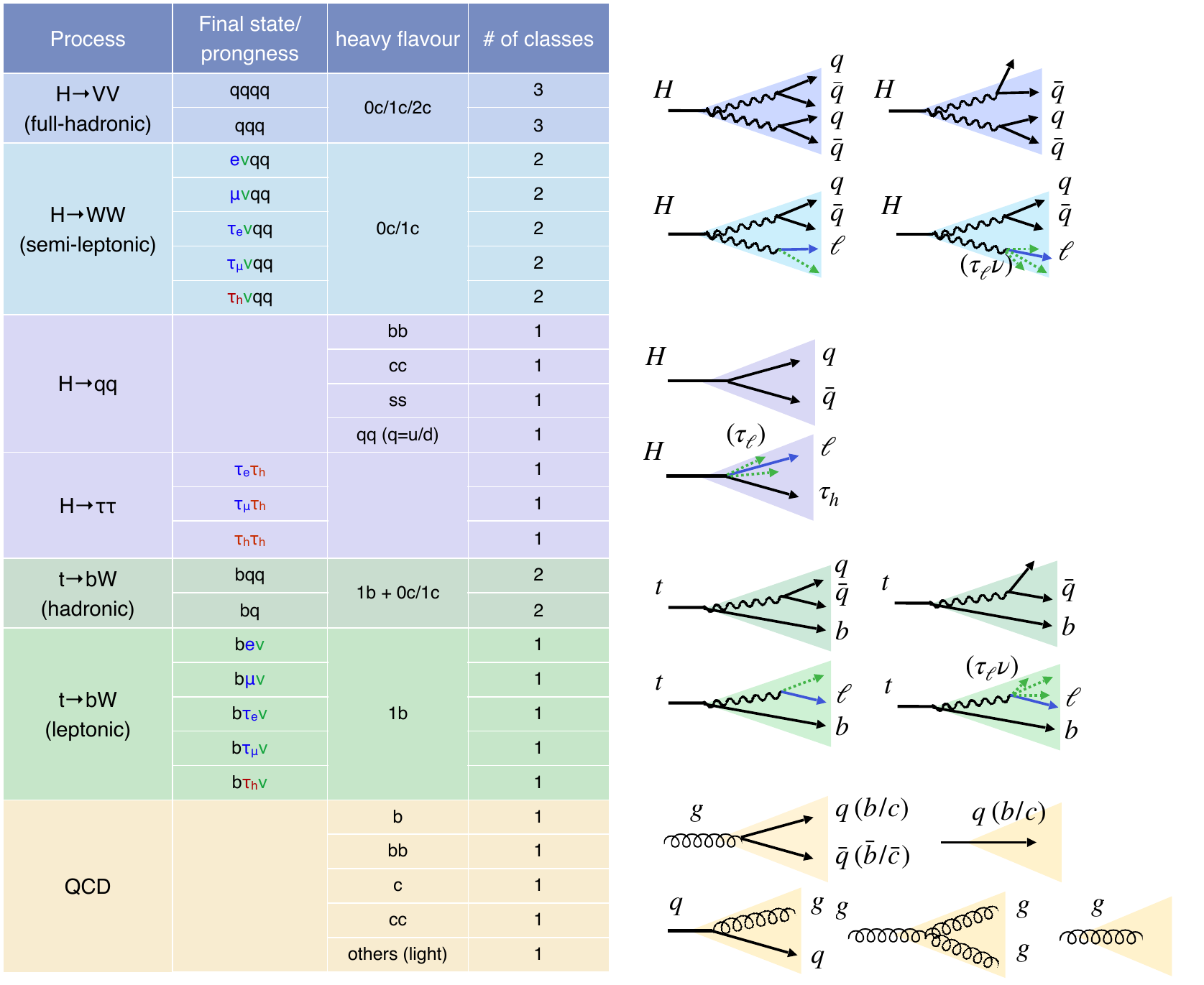}
    \includegraphics[width=0.22\linewidth,trim={0 0 0 15mm},clip]{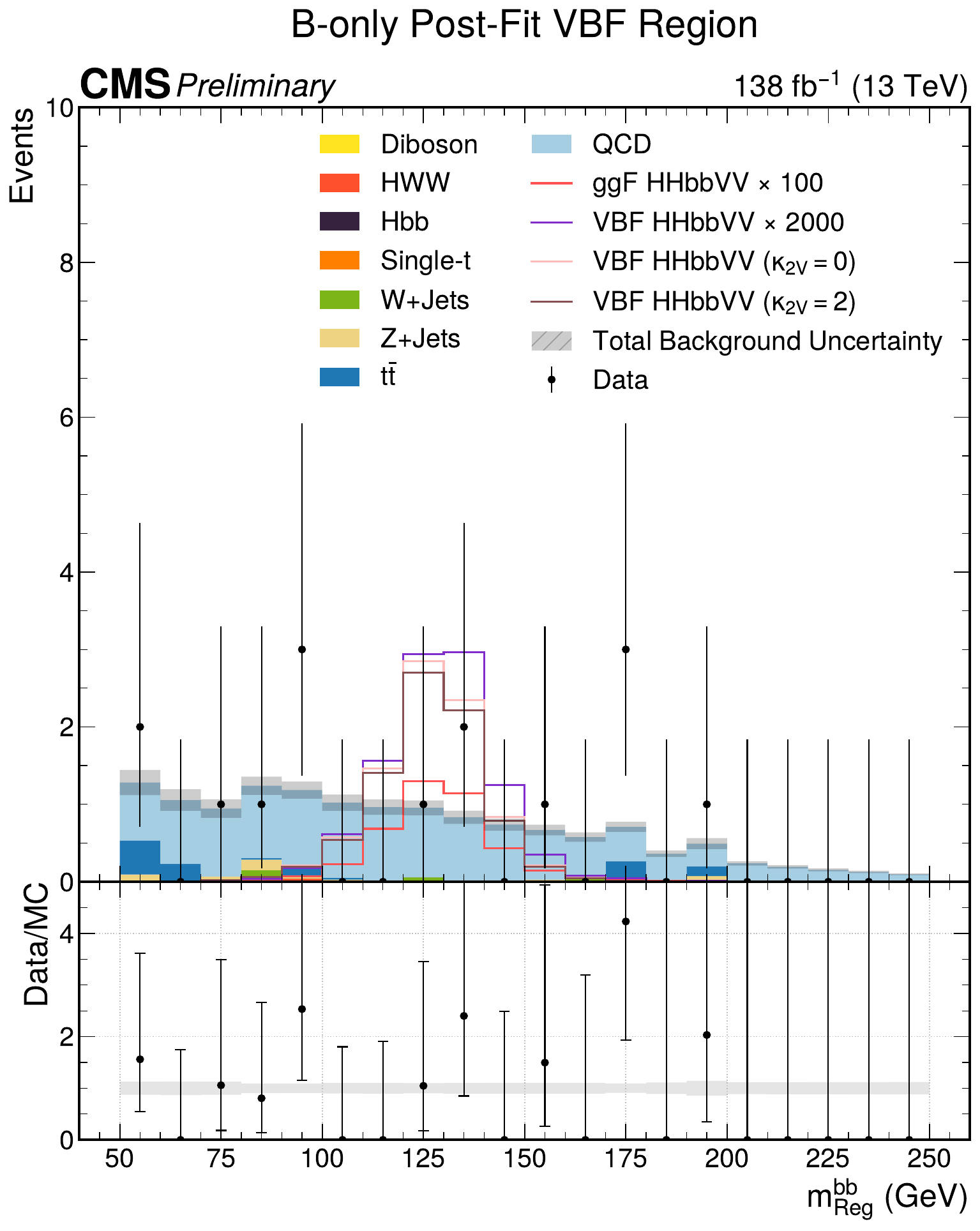}
    \includegraphics[width=0.34\linewidth]{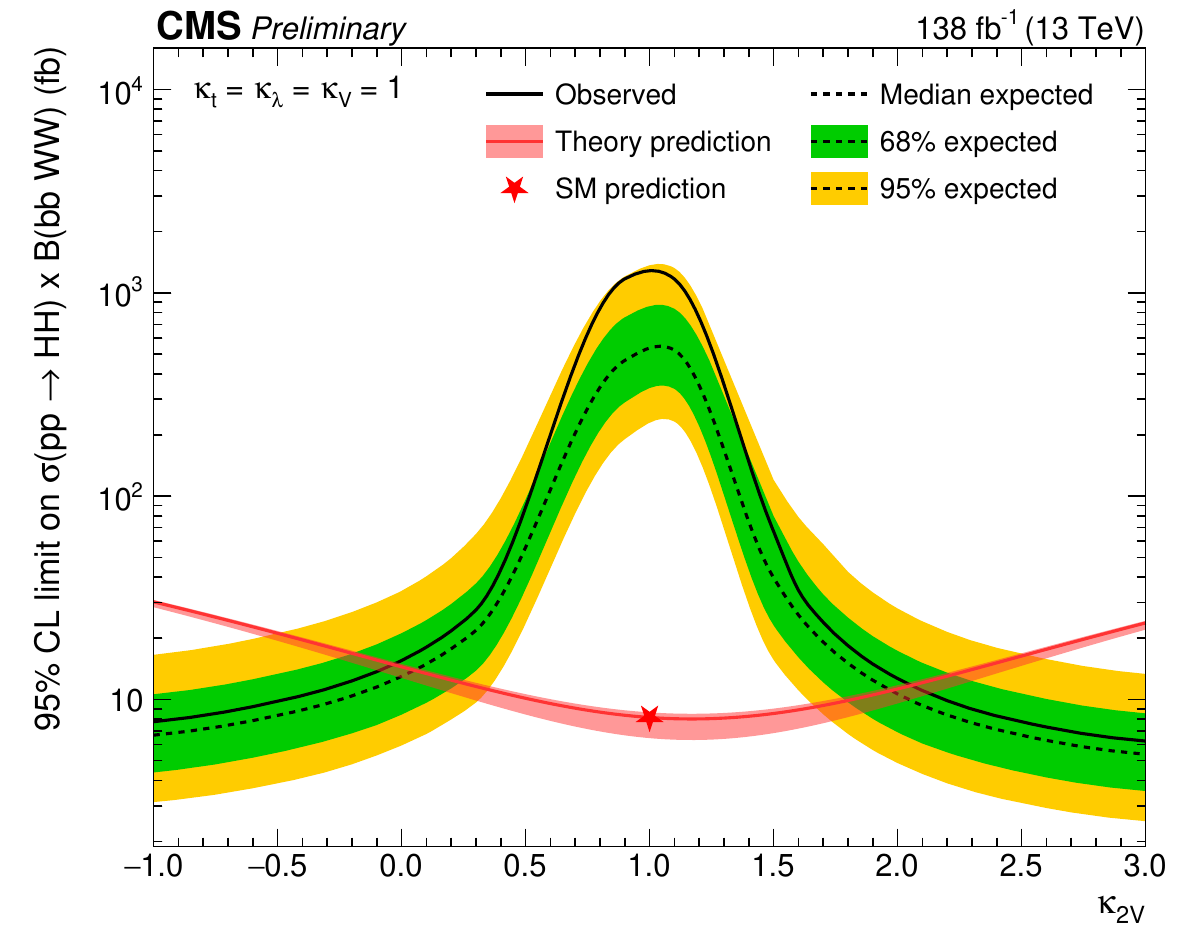}
    \caption{Full set of training jet classes for GloParT (left).
    Distributions of the $b\overline{b}$-candidate jet mass in the VBF signal region (center).
    Upper limits on the inclusive $HH$ production cross section as a function of $\kappa_{2V}$ (right).}
    \label{fig:hhbbvv}
\end{figure}

Building on the success of ParticleNet in CMS for large-radius jet tagging, which focused on discriminating $X(qq)$, $X(cc)$, $X(bb)$, and QCD jets, a new algorithm, dubbed Global ParT (GloParT)~\cite{CMS-PAS-HIG-23-012} is trained on a much larger set of classes, including all-hadronic and semileptonic $X(VV)$ decays, as shown in Fig.~\ref{fig:hhbbvv} (left).
This algorithm is based on ParT~\cite{Qu:2022mxj}, which leverages a learned ``attention'' and pairwise features to give more weight to certain particles, and disregard others, in order to infer the origin of jets.
Since this algorithm is trained in Monte Carlo (MC) simulation, it can be a challenge to calibrate in data when there is no SM analogue for the signal, such as the $H\to VV \to 4q$, that we can isolate.
However, a new calibration technique uses the Lund jet plane~\cite{CMS-DP-2023-046}.
Effectively, this technique measures data-to-simulation corrections per quark subjet, allowing the appropriate corrections to be determined with a readily available data sample of $W(qq)$ jets.

This tagger enables a new search for nonresonant, boosted $HH\to b\overline{b}VV\to b\overline{b}4q$~\cite{CMS-PAS-HIG-23-012}.
The vector boson fusion (VBF) $HH$ production mode is especially sensitive to a coupling modifier of the two-Higgs-boson-two-vector-boson ($HHVV$) interaction in the SM, known as $\kappa_{2V}$.
If $\kappa_{2V}$ differs from the SM value of 1, the differential VBF $HH$ production cross section dramatically increases at high $m_{HH}$, resulting in enhanced $HH$ production.
Figure~\ref{fig:hhbbvv} (center) displays the $b\overline{b}$-candidate jet mass after a tight GloParT $H\to 4q$ tagger and VBF-like selection, where the BSM signal would be clearly visible.
The search provides the second-best constraint on this coupling in the CMS experiment, as illustrated in Fig.~\ref{fig:hhbbvv} (right).

\begin{figure}[htpb]
    \centering
    \includegraphics[width=0.4\linewidth]{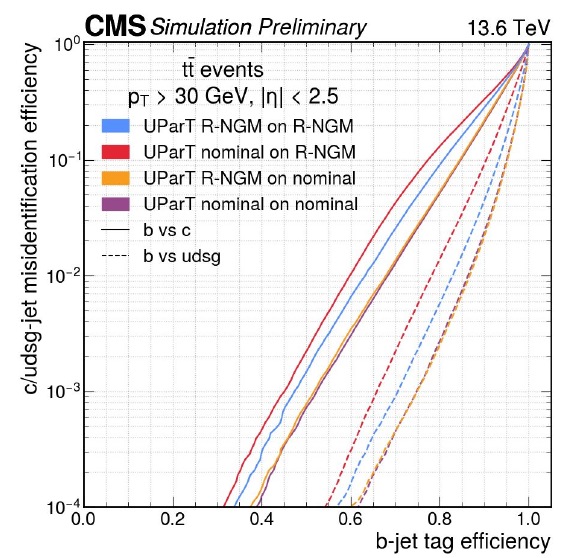}
    \includegraphics[width=0.37\linewidth]{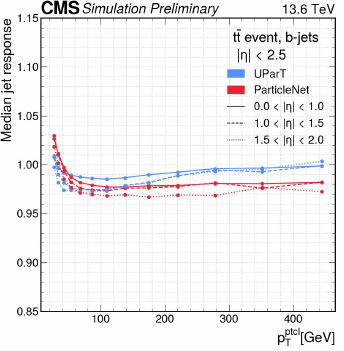}
    \caption{
    The ROC curves for UParT trained with R-NGM or nominal samples and evaluated with R-NGM or nominal samples (left).
    Median of the raw regressed jet energy response for UParT and ParticleNet (right). 
    \vspace{-1.5em}}
    \label{fig:upart}
\end{figure}

For small-radius jets in CMS, a similar algorithm has been developed that unifies heavy flavor tagging, hadronic tau tagging, jet energy regression, and jet energy resolution estimation, called unified particle transformer (UParT)~\cite{CMS-DP-2024-066}.
An additional novel aspect is the use of a rectified normed gradient method (R-NGM) adversarial training to improve model robustness.
In this strategy, the inputs are perturbed by adding a term proportional to the absolute value of the gradient of the loss function, which maximally disrupts the performance.
The trained model can then generalize better when faced with perturbed or mismodeled inputs.
This is illustrated in the receiver operating characteristic (ROC) curves shown in Fig.~\ref{fig:upart} (left), where UParT trained with R-NGM shows a substantial improvement compared to the nominal training when evaluated with the perturbed inputs, while preserving a similar performance when evaluated with the nominal inputs.
Figure~\ref{fig:upart} (right) also demonstrates the superior jet energy regression performance of UParT over a previous algorithm based on ParticleNet.
%Calibration is in progress for the 2022 and 2023 data~\cite{CMS-DP-2024-064}.

\begin{figure}[htpb]
    \centering
    \includegraphics[width=0.5\linewidth]{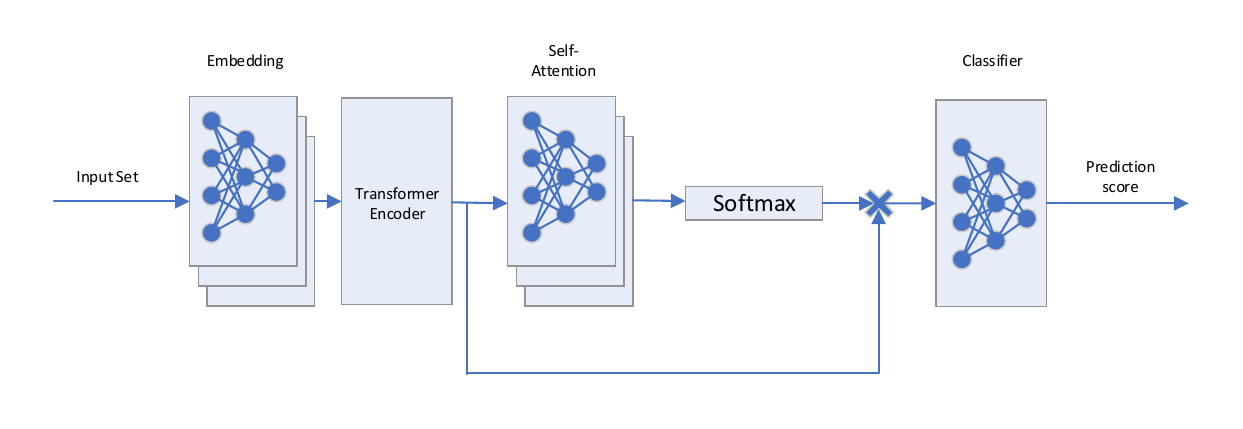}
    \includegraphics[width=0.4\linewidth]{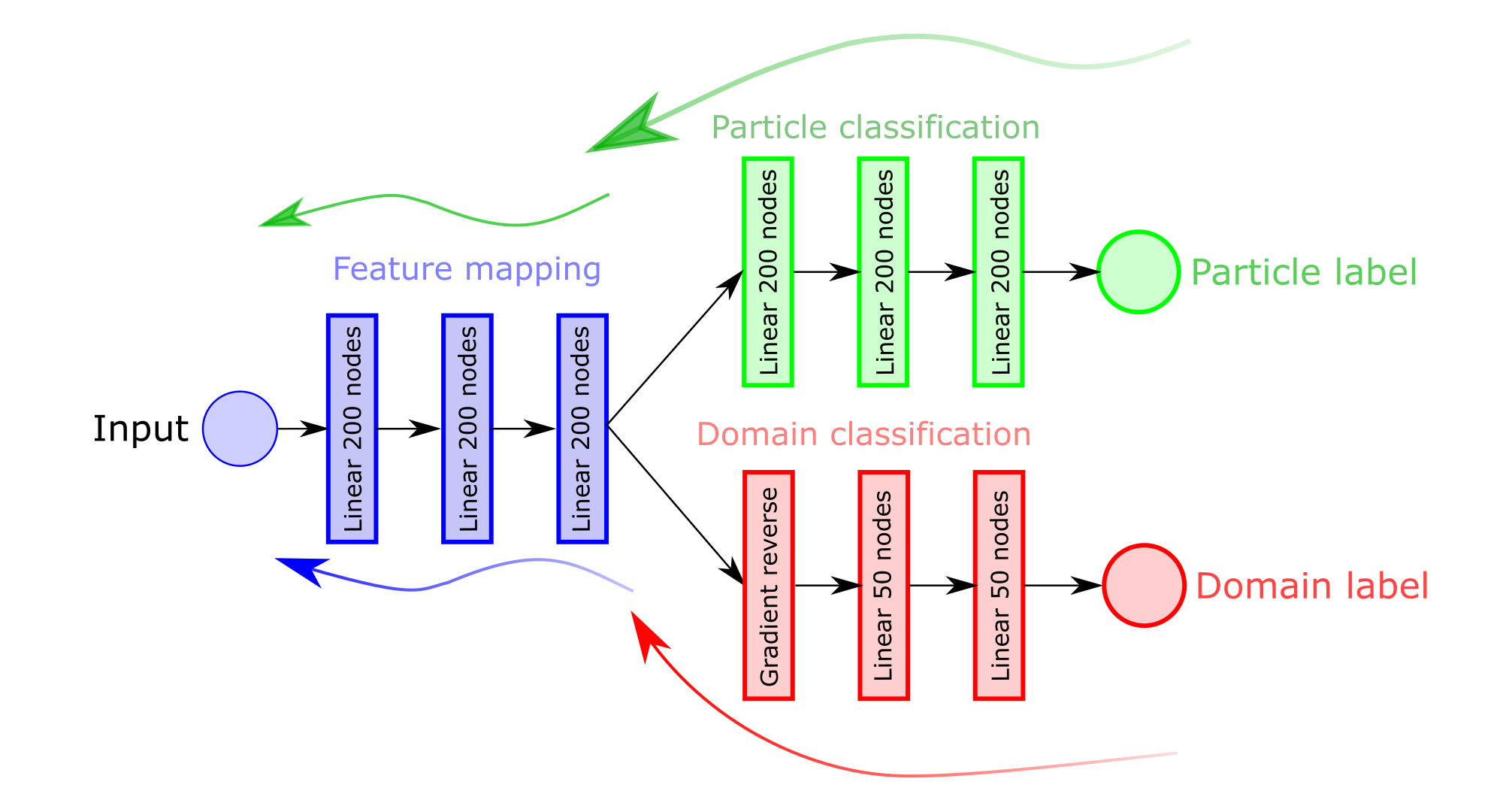}
    \caption{Transformer architecture for particle identification in ALICE (left).
    Layered blocks are applied separately to each vector in a set.
    Single blocks are applied to their input as a whole.
    Domain-adversarial neural network training setup (right).
    }
    \label{fig:alice}
\end{figure}

In the ALICE experiment, transformers have been used for particle identification, which results in higher purity and efficiency than standard methods, even for data with missing values due to limited detector efficiency and acceptance~\cite{Kasak:2023hhr,Karwowska:2024xqy}.
The proposed model architecture is based on a transformer, as shown in Fig.~\ref{fig:alice} (left).
A domain-adversarial neural network approach consisting of three neural networks is also proposed to mitigate differences between the two ``domains'' of data and simulation.
In this setup, a featurizer maps the original input to domain-invariant features that are provided to the particle classifier and the domain classifier, which enforces the domain invariance of the extracted features through an adversarial training, as shown in Fig.~\ref{fig:alice} (right).

\begin{figure}[htpb]
    \centering
    \includegraphics[width=0.5\linewidth]{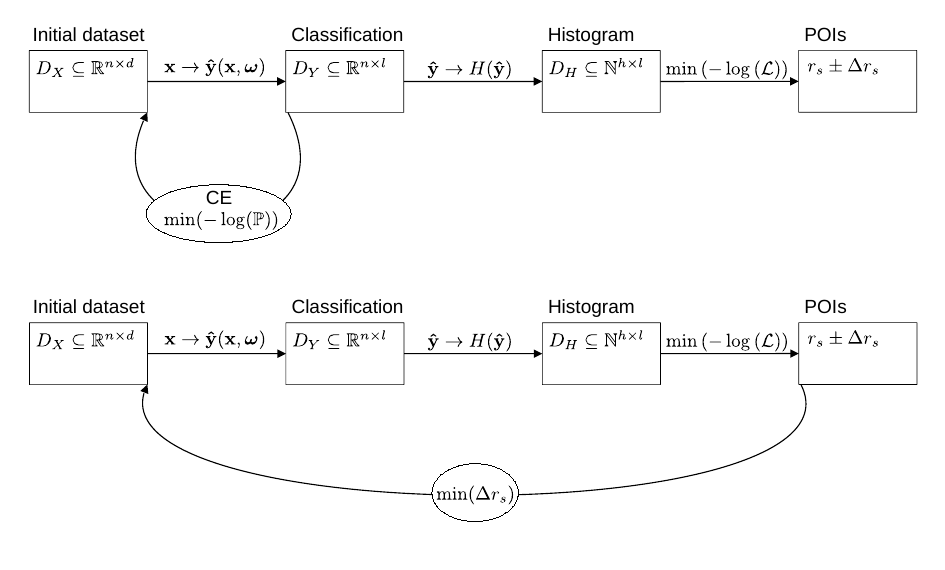}
    \includegraphics[width=0.2\linewidth]{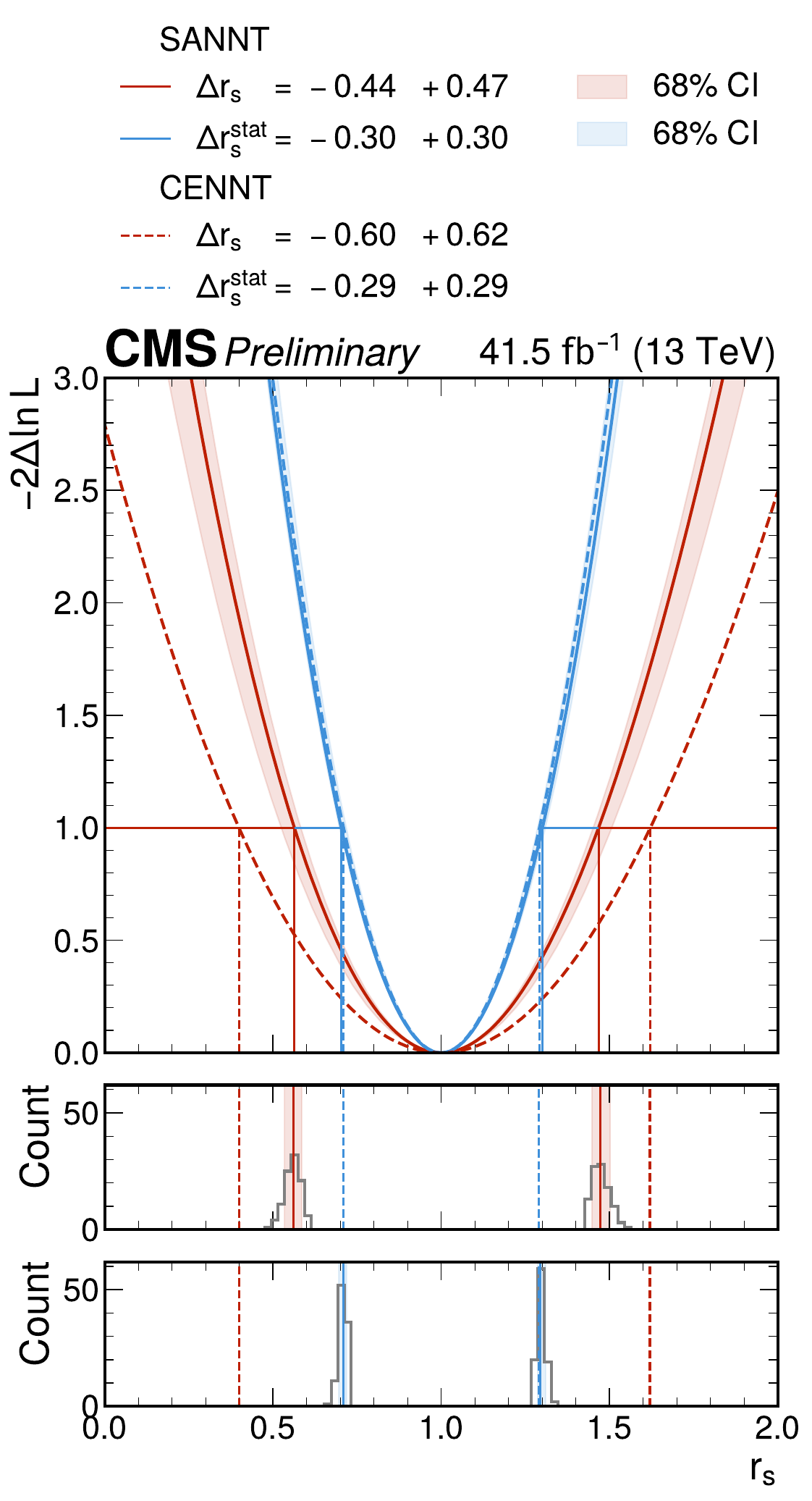}
    \caption{Flow chart of a CENNT (upper left) and SANNT (lower left).
    Negative log of the profile likelihood $-2\Delta \ln L$ as a function of $r_s$, taking into account (red) all and (blue) only the statistical uncertainties in $\Delta r_s$ (right).
    The results as obtained from CENNT (SANNT) are indicated by the dashed (solid) lines.\vspace{-1.5em}}
    \label{fig:sannt}
\end{figure}

Finally, a new systematic-aware neural network training (SANNT) has been developed in the context of a $H\to\tau\tau$ search in CMS~\cite{CMS-PAS-MLG-23-005}.
Traditionally, classifier neural networks are trained using the cross entropy loss function (cross entropy neural network training, or CENNT), which optimizes for signal versus background discrimination without considering systematic effects that may influence the ultimate figure of merit: the measurement uncertainty $\Delta r_s$ on a physics model's parameter of interest $r_s$, such as a signal strength.
Alternatively, by implementing the full analysis chain, including systematic uncertainties, in a differentiable way, it is possible to directly minimize $\delta r_s$ as the neural network loss function using gradient descent.
Flow charts of these two contrasting approaches are shown in Fig.~\ref{fig:sannt} (left).
A crucial ingredient is the choice of custom functions $\mathcal B_i$ to substitute for the gradients of the non-differentiable histogram operation $\mathbf{\hat y} \to H(\mathbf{\hat y})$.
Based on SANNT 
(CENNT), values of $\Delta r_s=^{+0.47}_{-0.44}$ ($\Delta r_s=^{+0.62}_{-0.60}$) for the $H\to\tau\tau$ signal strength are obtained, corresponding to a 25\% reduction in the measurement uncertainty for SANNT compared to CENNT as shown in Fig.~\ref{fig:sannt} (right).
For the first time in an analysis of this complexity, the significant gains of SANNT over CENNT are demonstrated.
\vspace{-1.5em}

\vspace{-.5em}
\section{Fast simulation}
\vspace{-.5em}

\begin{figure}[htpb]
    \centering
    \includegraphics[width=0.3\linewidth]{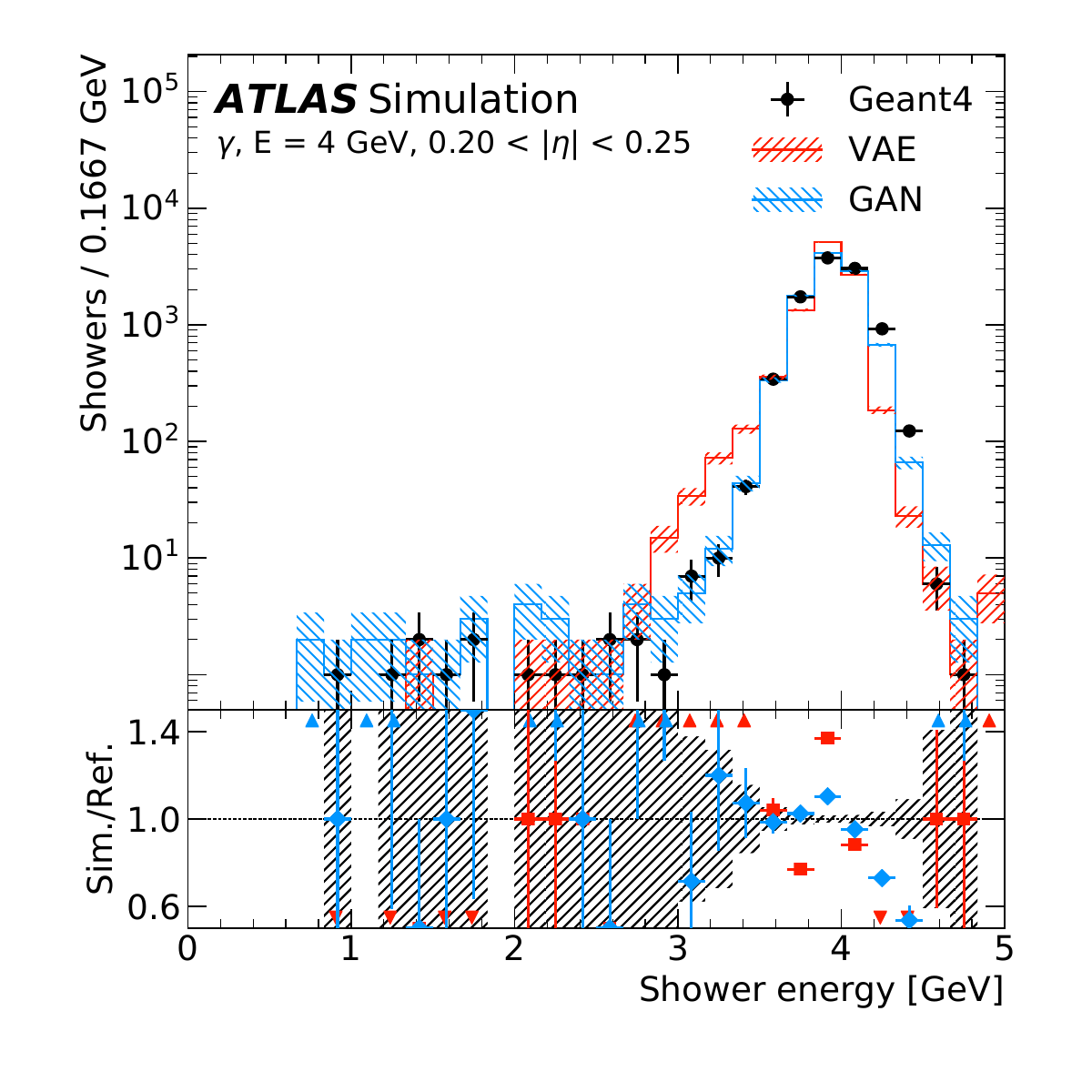}
    \includegraphics[width=0.3\linewidth]{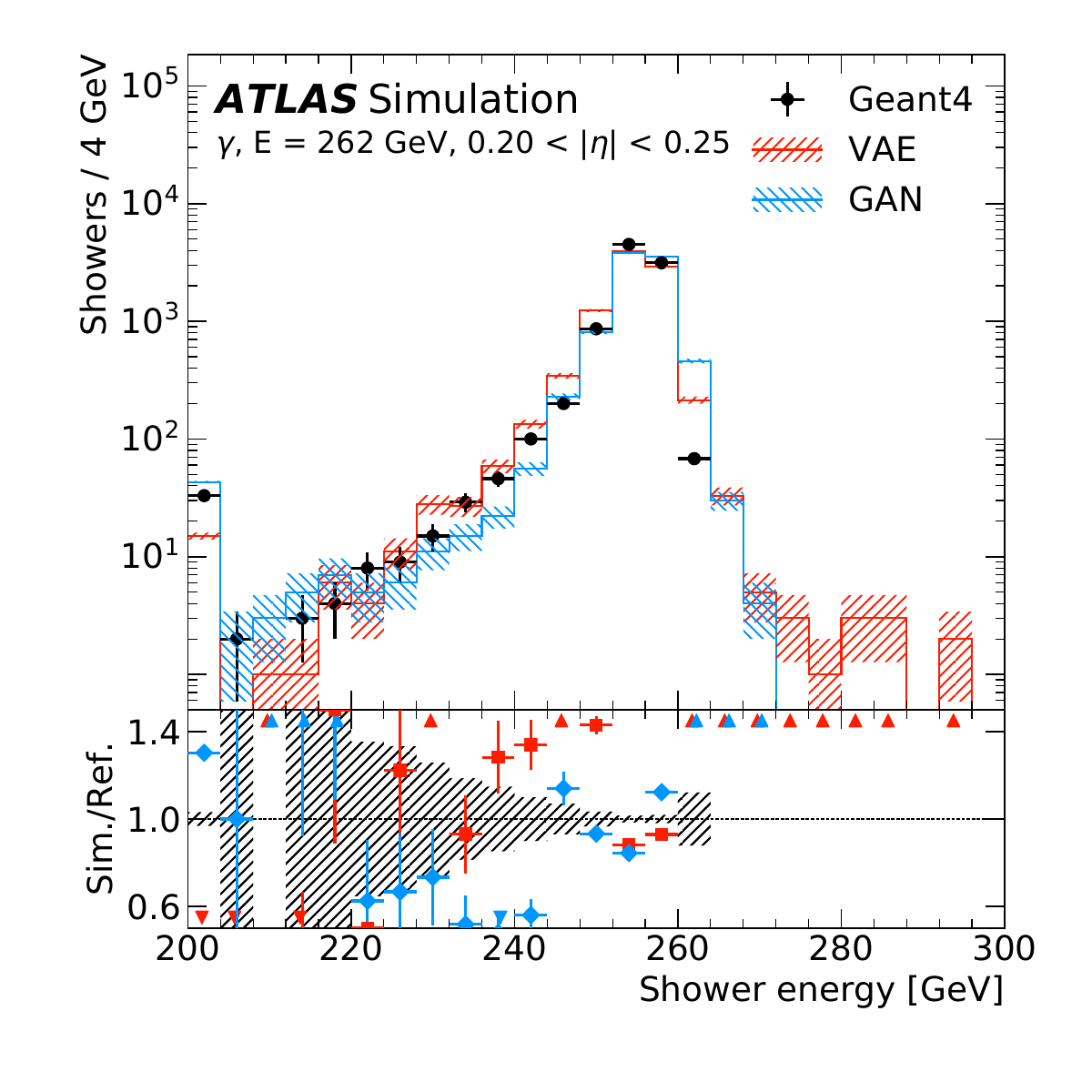}
    \includegraphics[width=0.24\linewidth]{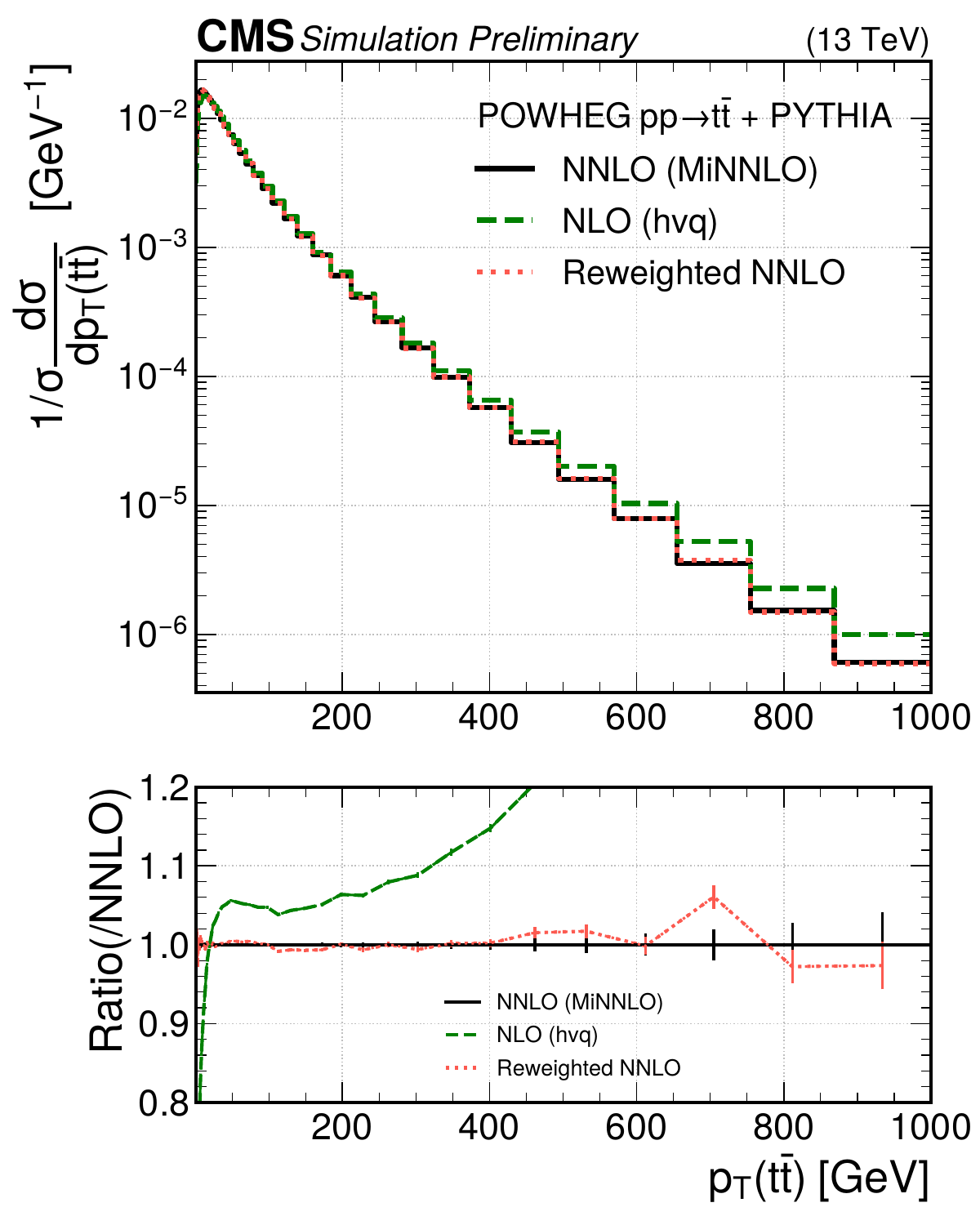}
    \caption{Total energy response of the calorimeter to photons with an energy of 4 GeV (left) and 262 GeV (center).
    The calorimeter response for the GEANT4 training data (black markers) are shown as reference points and compared with those from a VAE (solid red line) and a GAN (solid blue line).
    Distributions in $p_\mathrm{T}$ of the $t\overline{t}$ system obtained from simulations at next-to-next-to-leading-order (NNLO) accuracy (black solid lines), next-to-leading-order (NLO) accuracy (green dashed lines), and NLO reweighted to NNLO (red dotted lines) (right).
    \vspace{-1.5em}
    }
    \label{fig:fastcalogan_dctr}
\end{figure}

Generating large-scale, realistic simulated data samples is a key driver of the increased CPU needs for the high-luminosity LHC.
Machine learning methods can be used to ``short cut'' traditional simulation, producing simulated data samples of sufficient quality for analysis at a fraction of the CPU cost.
The ATLAS Collaboration has investigated the use of variational autoencoders (VAEs) and generative adversarial networks (GANs) to model the response of the ATLAS electromagnetic calorimeter to photons of various energies~\cite{ATLAS:2022jhk}.
Compared to a full detector simulation using \textsc{Geant4}, both VAEs and GANs are found to be capable of quickly simulating electromagnetic showers with correct total energies (Fig.~\ref{fig:fastcalogan_dctr}, left and center) and stochasticity, though the modeling of some shower shape distributions requires more refinement.
% This study demonstrates the potential of ML for fast calorimeter simulation in the future.

Another approach to reduce the computational costs of simulation is the use of a neural network to reweight from one simulated sample to another one with different physics model parameters or to match more accurate (higher order in perturbation theory) calculations. 
This method circumvents the need to simulate the detector response multiple times by reusing a single sample with different learned event weights.
The method relies on the \emph{likelihood ratio trick}, in which a classifier $f(x)$ trained to distinguish samples from two different probability distribution functions, $p_0(x)$ or $p_1(x)$, approximates the likelihood ratio $f(x)/(1-f(x)) \approx p_0(x)/p_1(x)$, which can be used as an event weight~\cite{Andreassen:2019nnm}.
The CMS Collaboration has used this method for several applications in top quark physics, including to reweight MC simulation to account for higher-order theory predictions, as shown in Fig.~\ref{fig:fastcalogan_dctr} (right)~\cite{CMS-PAS-MLG-24-001}.
% A particle flow network based on the parton kinematics and ID of the $t\overline{t}$ system is used to reweight the original sample and the reweighted sample agrees to within 2\% of the target distribution.

\vspace{-.5em}
\section{Unfolding}
\vspace{-.5em}

\begin{figure}[htpb]
    \centering
    \includegraphics[width=0.8\linewidth]{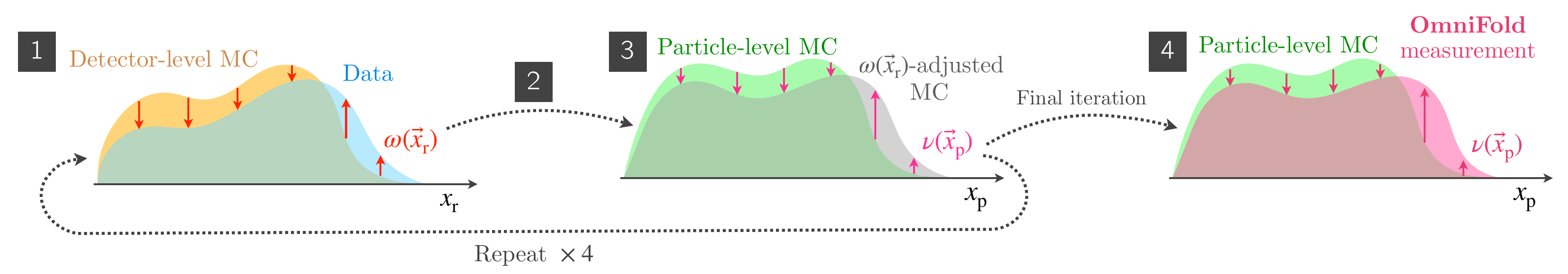}\\
    \includegraphics[width=0.35\linewidth]{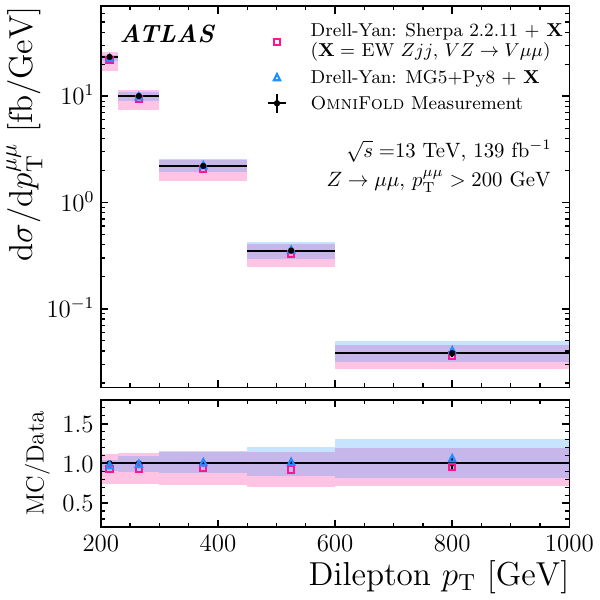}
    \includegraphics[width=0.4\linewidth]{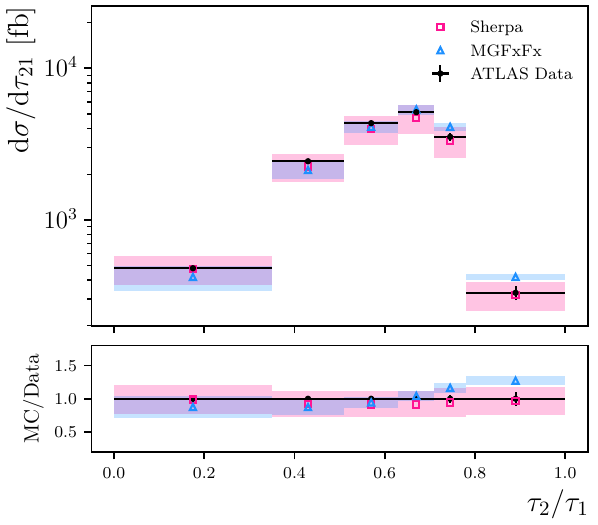}
    \caption{Illustration of the OmniFold method (upper).
    Measured differential cross sections compared with particle-level predictions from \textsc{Sherpa} and \textsc{MadGraph} for one of the 24 directly measured observables in Ref.~\cite{ATLAS:2024xxl}, the dilepton $p_\mathrm{T}$ (lower left), and another observable that can be derived with the released dataset the leading jet substructure variable $\tau_{2}/\tau_{1}$ (lower right).}
    \label{fig:omnifold}
\end{figure}

Machine learning methods can also be used to ``unfold'' detector effects and measure the differential distributions of physics observables at the particle level.
The ATLAS Collaboration has employed the OmniFold method~\cite{Andreassen:2019cjw} to produce a simultaneous measurement of 24 $Z$+jets observables~\cite{ATLAS:2024xxl}.
In this method, the detector-level MC simulation is first corrected by a learned weighting function $\omega(\vec x_r)$ to match data.
Then, the particle-level MC simulation is corrected by another learned weighting function $\nu(\vec x_p)$ to match the $\omega(\vec x_r)$-weighted MC simulation.
The method is iterated four more times, to achieve $\nu(\vec x_p)$-weighted MC events whose event yields and kinematics match those observed in data.
Unlike previous fiducial differential cross section measurements, the result is presented as an unbinned dataset of weighted particle-level events~\cite{atlas_collaboration_2024_11507450}, allowing for new observables to be constructed from the 24 measured observables as shown in Fig.~\ref{fig:omnifold} (lower right).

\vspace{-.5em}
\section{Anomaly detection}
\vspace{-.5em}

If the form of new physics is not known, it can be difficult if not impossible to design a search to target it.
Anomaly detection methods offer the promise of sensitivity to a broad range of new physics signatures, enabling model-agnostic searches.
There are several varieties of ML-based anomaly detection techniques, ranging from unsupervised methods, in which no labels are known during training, to weakly supervised methods, in which labels are partially known.
A classic example of an unsupervised method is a (variational) autoencoder, or (V)AE, that compresses input data then attempts to reconstruct it.
An anomaly score can then be built based on the distance between the input and the output---the assumption being that those examples that are misreconstructed are outside of the training domain and thus anomalous.

\begin{figure}[htpb]
    \centering
    \includegraphics[width=0.35\linewidth]{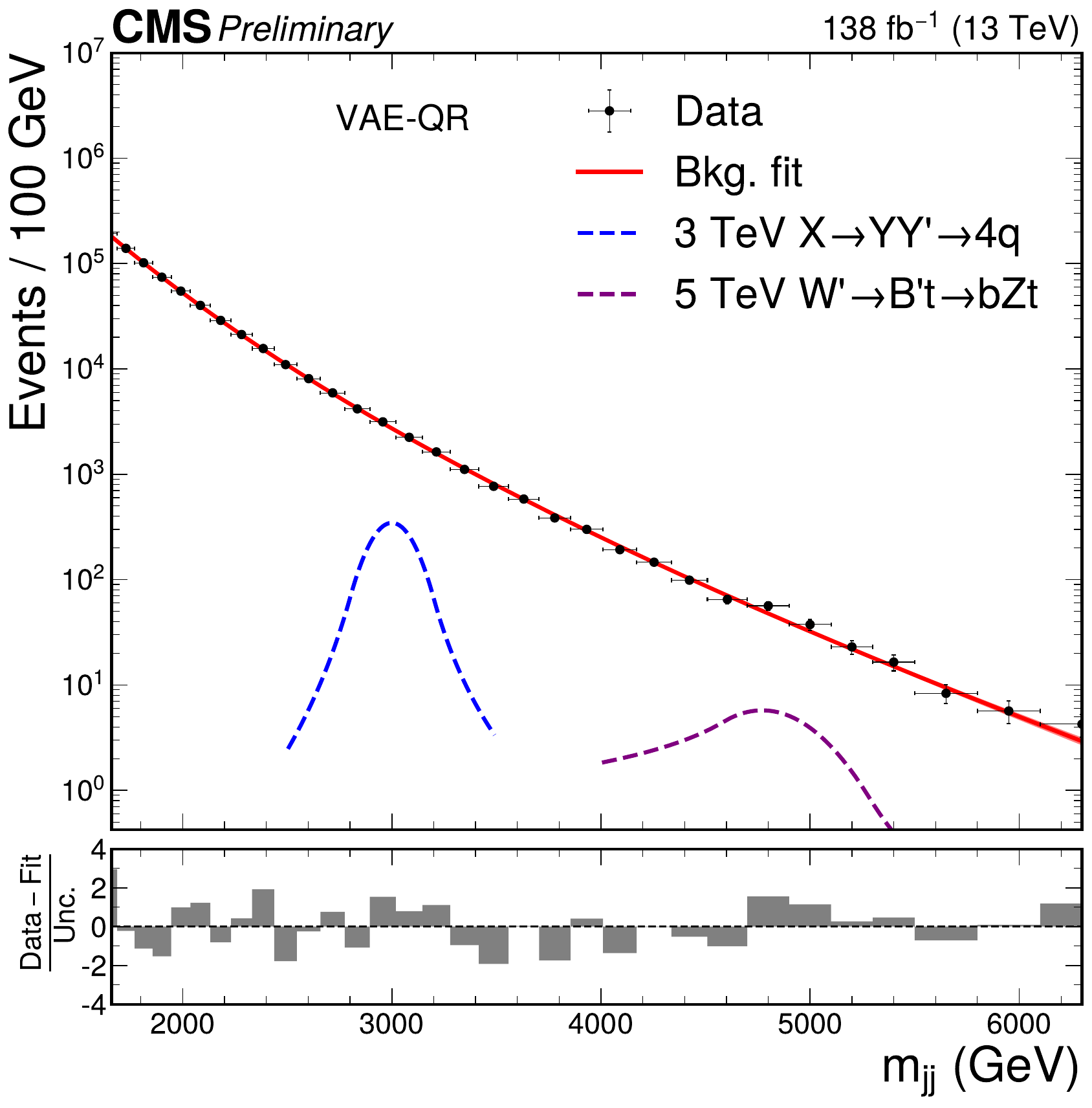}    
    \includegraphics[width=0.33\linewidth]{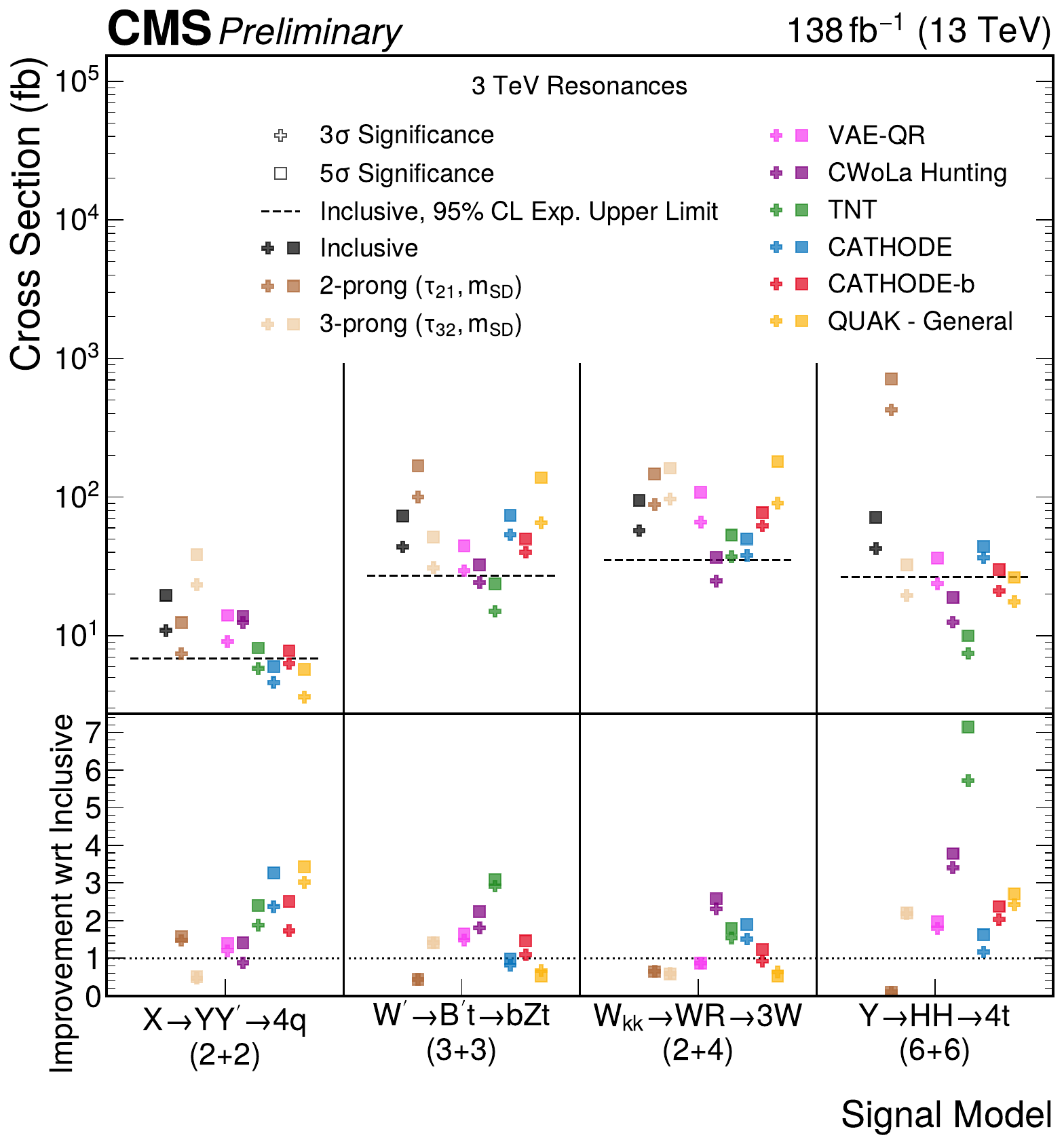}
    \caption{The dijet invariant mass spectrum and resulting background fit to the data for the VAE-QR method, along with example signals (left). 
    The cross section that would lead to an expected $3\sigma$ ($5\sigma$) excess is shown as a cross (square) marker for the six anomaly detection methods (six colors), an inclusive dijet search (black), and traditional substructure cuts targeting two-pronged (dark brown) or three-pronged decays (tan) (right).
    \vspace{-0.5em}
 }
    \label{fig:case}
\end{figure}

In the CMS experiment, five different anomaly detection techniques are used to search for new physics in dijet events~\cite{CMS-PAS-EXO-22-026}, including a VAE with quantile regression (VAE-QR) to ensure a constant efficiency as a function of the invariant mass of the dijet.
Figure~\ref{fig:case} (left) shows the dijet invariant mass spectrum and resulting background fit to the data for the VAE-QR method, along with example signals. 
The methods reduce the cross section needed for a $5\sigma$ discovery compared to the inclusive dijet search by up to factor of 7, as shown in Fig.~\ref{fig:case} (right).

\begin{figure}[htpb]
    \centering
    \includegraphics[width=0.35\linewidth]{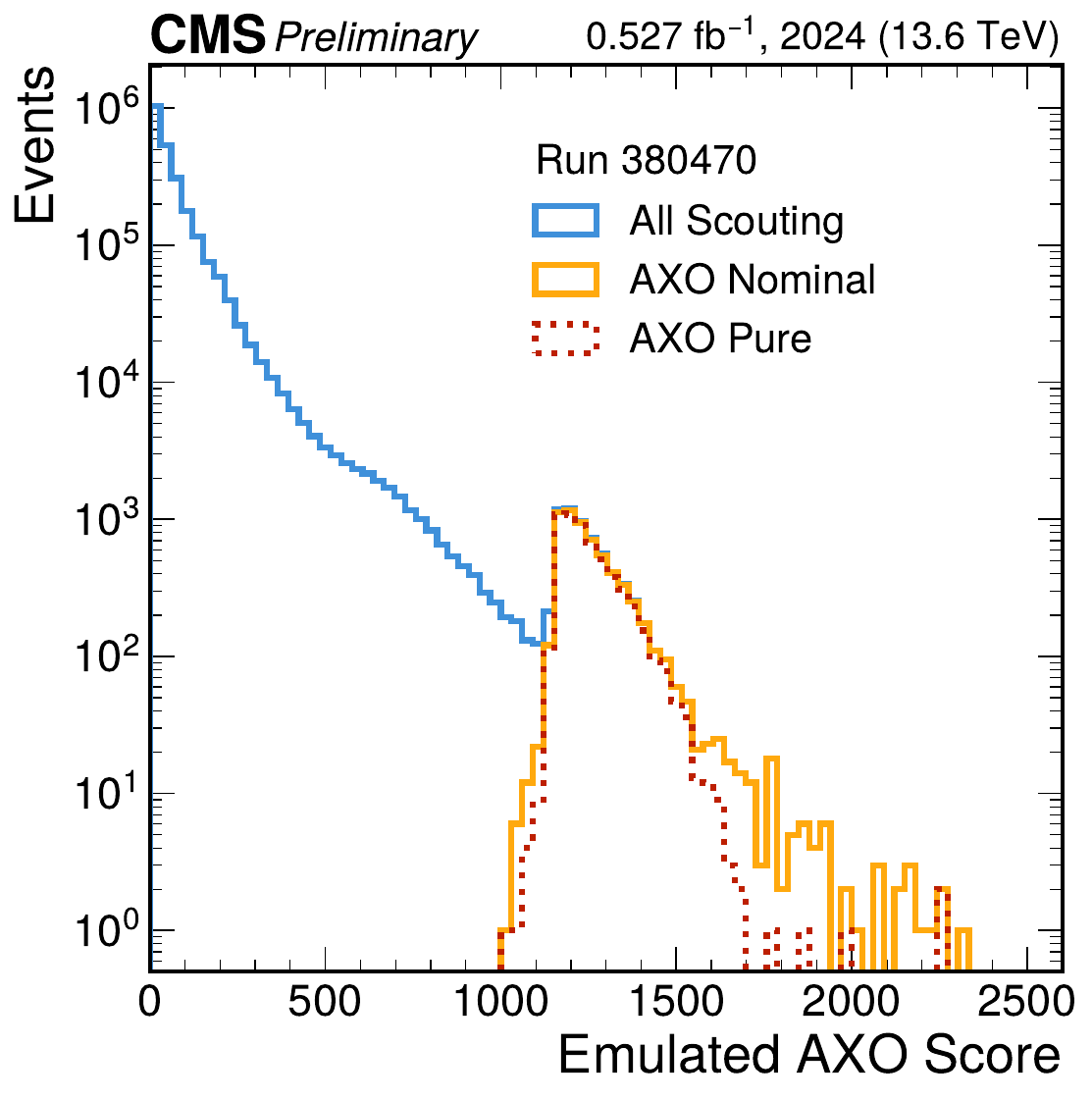} 
    \includegraphics[width=0.35\linewidth]{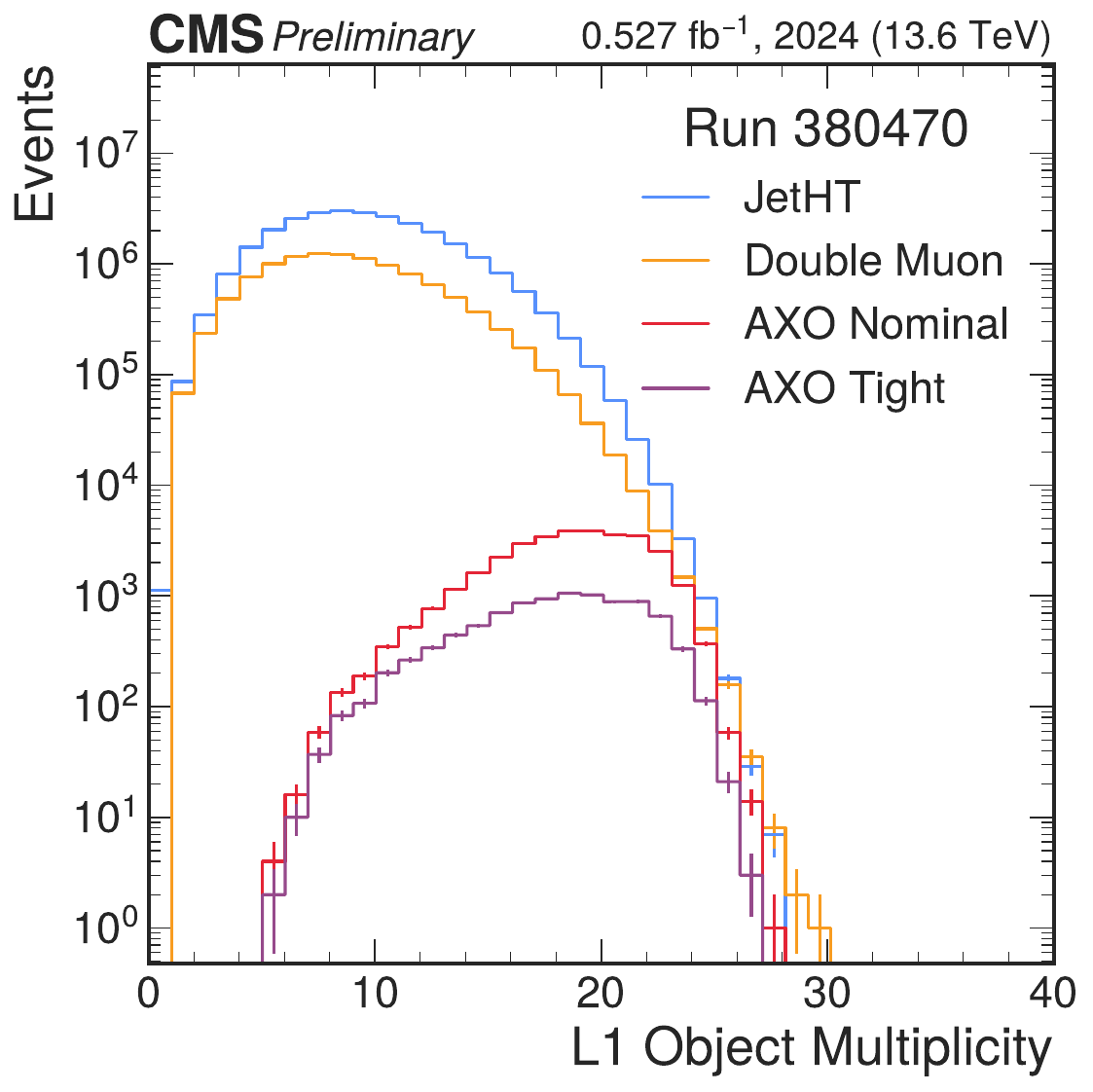}
    \caption{AXOL1TL score distributions for the HLT scouting partial dataset for CMS Run 380470 on May 7, 2024. Scores are shown for all events triggered by the nominal seed, and those only triggered by the AXOL1TL triggers and no other level-1 triggers (pure).    
    Level-1 trigger object multiplicity distributions for jet $H_\mathrm{T}$, double muon, and AXOL1TL triggers (right).}
    \label{fig:axol1tl}
\end{figure}

Anomaly detection algorithms can be applied to select events directly at the trigger level.
Such an algorithm, called AXOL1TL, has been developed for the CMS level-1 global trigger, consisting of a VAE trained on zero bias data events using the kinematic information of up to 10 jets, 4 muons, 4 electrons, and the missing transverse momentum in each event as input~\cite{CMS-DP-2023-079,CMS-DP-2024-059}.
The encoder is compiled into firmware for an FPGA with the hls4ml package~\cite{Duarte:2018ite}, and an anomaly score is calculated based on the latent representation.
%The algorithm has been deployed for data collection in 2024.
Figure~\ref{fig:axol1tl} (left) shows the score distributions during 2024 data collection.
In particular, the pure contribution shows that AXOLT1TL selects unique events relative to existing level-1 trigger.
However, there is also a preference for high-multiplicity events (Fig.~\ref{fig:axol1tl}, right), which leads to an undesirable pileup dependence.

\begin{figure}[htpb]
    \centering
    \includegraphics[width=0.4\linewidth]{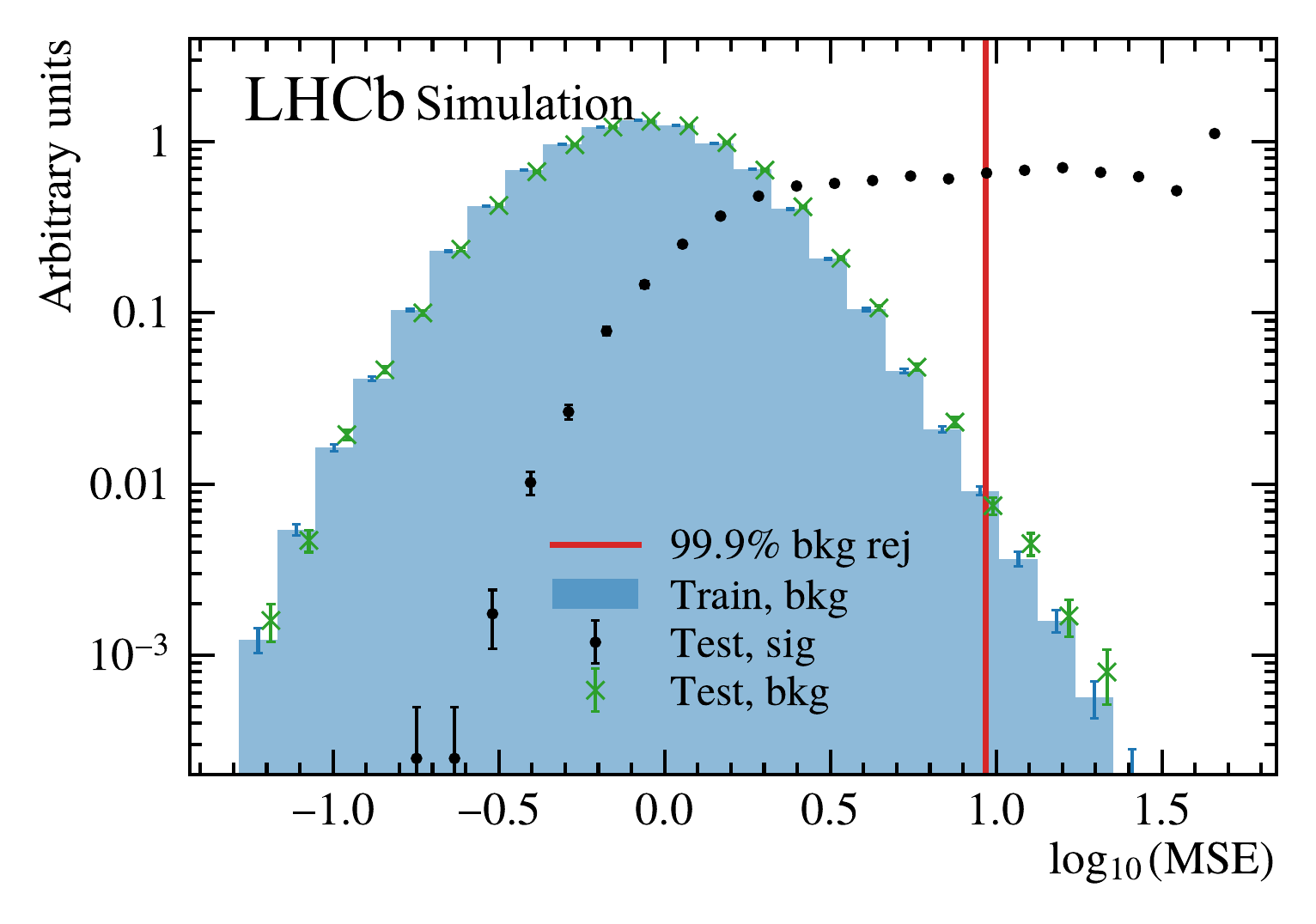}
    \includegraphics[width=0.4\linewidth]{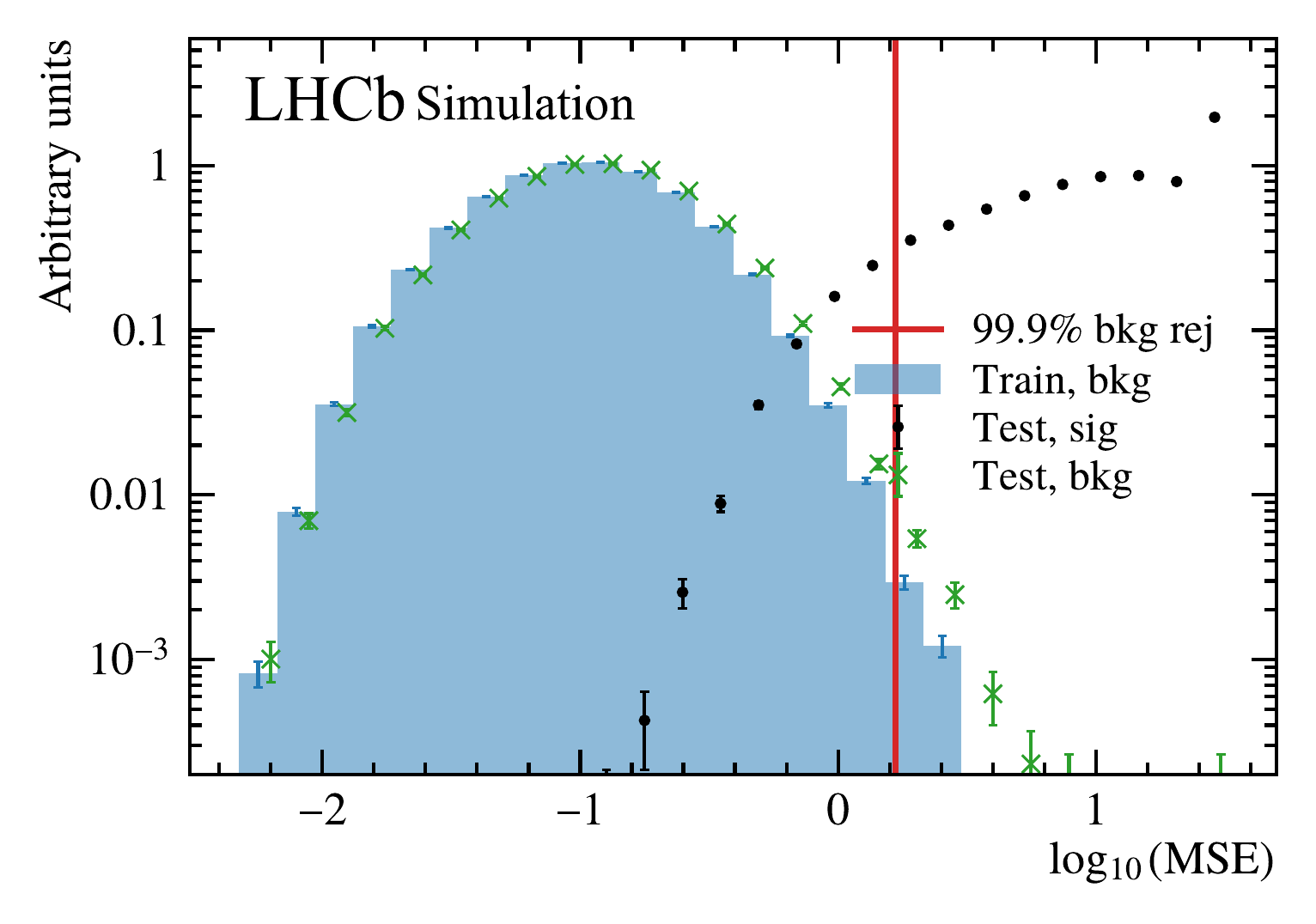}
    \caption{Distribution of the predictions for the background and signal samples, using the autoencoder (left) and the normalized autocndoder (right).
    The signal consists of long-lived axions produced in a Higgs boson decay $H \to AA$, decaying to tau leptons $A \to \tau\tau$, where each tau lepton decays to three pions.}
    \label{fig:normalized_autoencoder}
\end{figure}

At times, autoencoders can be too good at reconstructing signal, meaning that signal is not flagged as anomalous due to a low reconstruction error.
The normalized autoencoder approach~\cite{Dillon:2022mkq} is designed to mitigate this issue by aligning the low reconstruction error phase space with the background phase space.
The LHCb experiment has implemented this approach to trigger on long-lived particles decaying in the muon chambers~\cite{LHCB-FIGURE-2024-015}.
The idea is to promote the autoencoder to an energy-based model that models the data as $p_\theta(x)\propto \exp\left[\mathrm{MSE}(x, \mathrm{AE}(x))\right]$; the parameters of the autoencoder are then obtained by minimizing the negative log likelihood of the data $-\ln p_\theta(x)$.
In practice, this is approximately minimized through a clever reframing using Markov chain Monte Carlo to sample from $p_\theta(x)$.
With this change in training, the normalized autoencoder can capture more anomalous signals, as shown in Fig.~\ref{fig:normalized_autoencoder}

\vspace{-.5em}
\section{Summary and outlook}
\vspace{-.5em}

There is a dizzying array of ML opportunities, innovations, and applications in LHC experiments, which directly impact physics results.
Physicists are concerned not only with performance but also robustness, interpretability, and insensitivity to modeling uncertainties.
The use cases of ML extend beyond classification to include simulation, unfolding, anomaly detection, and more.
ML methods have enabled new searches and measurements that were previously impossible.

\vspace{-.5em}

\bibliographystyle{cms_unsrt}
\bibliography{skeleton}

\end{document}